\documentclass[prl, twocolumn, amsmath, amssymb]{revtex4}
\usepackage{graphicx}
\usepackage{amssymb}
\usepackage{color}

\input epsf

\definecolor{my_red}{rgb}{1,0.2,0.2}
\definecolor{my_green}{rgb}{0,1,0}
\definecolor{my_blue}{rgb}{0,0,0.4}
\definecolor{my_cyan}{rgb}{0.4,1,1}

\begin{document}

\title{Fourier relationship between angular position and optical orbital angular momentum}

\author{Eric Yao$^1$, Sonja Franke-Arnold$^1$, Johannes Courtial$^1$, Stephen Barnett$^2$ and Miles Padgett$^1$}

\affiliation{$^1$Department of Physics and Astronomy, University of Glasgow, Glasgow G12 8QQ, Scotland \\
$^2$ Department of Physics, University of Strathclyde, Glasgow G4 0NG, Scotland}


\begin{abstract}

\noindent We demonstrate the Fourier relationship between angular position and angular momentum for a light mode.  In particular we measure the distribution of orbital angular momentum states of light that has passed through an aperture and verify that the orbital angular momentum distribution is given by the complex Fourier-transform of the aperture function.  We use spatial light modulators, configured as diffractive optical components, to define the initial orbital angular momentum state of the beam, set the defining aperture, and measure the angular momentum spread of the resulting beam.  These measurements clearly confirm the Fourier relationship between angular momentum and angular position, even at light intensities corresponding to the single photon level.

\end{abstract}

\maketitle

{\bf Introduction.}
The spin angular momentum of a light beam is manifest as circular polarisation and can be attributed to the helicity of
individual photons. By contrast, the orbital angular momentum of a light beam is manifest
in $\ell$ intertwined helical phase fronts, with an azimuthal phase term $\exp(i\ell \phi) $ ,  which carry an associated orbital angular momentum (OAM) of $\ell\hbar$ per photon \cite{Allen92}.
Both the spin and orbital angular momentum of light can be transferred to solid objects, causing them to rotate about their own axis or around the beam axis respectively \cite{Neil}. Quantized spin and OAM have been measured for single photons \cite{Mair}. Just like linear momentum and linear position, angular momentum and angular position are related by a Fourier relationship \cite{quantum},  linking the standard deviations of the measurements.  This is a purely
classical phenomenon, but the Fourier relation also holds for quantum observables and in the quantum regime the Fourier relation is associated with the Heisenberg uncertainty principle.  Although this concept forms the basis of various calculations \cite{Allen, OAM} and experiments \cite{experiments} its validity has never been directly tested.  Here we present measurements that test the Fourier relation between the orbital angular momentum of light and its azimuthal probability (i.e.~intensity) distribution.

{\bf Fourier Conjugate Pairs.}
Linear momentum and position are both unbounded and continuous variables of a physical system and are related by
a continuous Fourier transform. For angular momentum and angular position the $2 \pi$ periodic nature of the angle
variable means that the relationship is a Fourier-series leading to discrete values of the angular momentum.  Assuming a
Fourier relationship between the distribution of angular momenta, $\psi_\ell$, and the angular distribution,
$\Psi(\phi)$, we can express one observable as the generating function of the other \cite{Franke-Arnold},
\begin{eqnarray}\label{1}
\psi_\ell&=&\frac{1}{\sqrt{2\pi}}\int_{-\pi}^{+\pi} d \phi\Psi(\phi)\exp(-i\ell\phi),\\ \label{2}
\Psi(\phi)&=&\frac{1}{\sqrt{2\pi}}\sum_{\ell=-\infty}^{+\infty}\psi_\ell\exp(i\ell\phi).
\end{eqnarray}
When light passes through an aperture or mask with an angular dependance given by $\Psi_{\rm Mask}(\phi)$  its phase and/or intensity profile is modified such that
\begin{eqnarray}\label{3}
\Psi_{\rm Transmitted}(\phi)=\Psi_{\rm Incident}(\phi) \times \Psi_{\rm Mask}(\phi),
\end{eqnarray}
\noindent where for simplicity, we have omitted the normalisation factor.  If the incident light is in a pure OAM state, defined by a single value of $ \ell$, this simplifies to
\begin{eqnarray}\label{4}
\Psi_{\rm Transmitted}(\phi)=\exp(i \ell \phi) \times \Psi_{\rm Mask}(\phi).
\end{eqnarray}
\noindent Note that as with the light beam in (\ref{2}), the complex transmission function of the mask can be expressed in terms of its angular harmonics with Fourier coefficients $A_{n}$,
\begin{eqnarray}\label{5}
\Psi_{\rm Transmitted}(\phi) = \sum_{n=-\infty}^{+\infty}A_{n} \exp(i (\ell+n) \phi),
\end{eqnarray}
\noindent where  $\sum_{n=-\infty}^{+\infty}|A_{n}|^2$ is the total intensity transmission of the mask. This means that upon transmission, each OAM component of the incident light acquires OAM sidebands shifted by $\delta\ell=n$, where the amplitude of each component is given by the corresponding Fourier coefficient of the mask,
\begin{eqnarray}\label{6}
\psi_{\delta\ell}=A_{n=\delta\ell}.
\end{eqnarray}
In the experiments presented here we have used hard-edge aperture segments of width $\Theta$, i.~e.~$\Psi_{\rm
Mask}(\phi)=1$ for $-\Theta/2<\phi \leq \Theta/2$ and $0$  elsewhere.  A single-segment mask can be expressed in
terms of its Fourier coefficients as
\begin{eqnarray}\label{7}
\Psi_{\rm Mask}(\phi) =\frac{\Theta}{2\pi} \sum_{n=-\infty}^{+\infty}{\rm sinc} \left(\frac{n \Theta}{2}\right)\exp(i n \phi),
\end{eqnarray}
hence giving OAM sidebands with amplitudes
\begin{eqnarray}\label{8}
\psi_{\delta\ell}=A_{n=\delta\ell}= \frac{\Theta}{2\pi}  {\rm sinc} \left(\frac{\delta\ell \Theta}{2}\right).
\end{eqnarray}
More generally, any azimuthal intensity distribution with $m-$fold symmetry only has angular harmonics at multiples of $m$.  Extending the design of the masks to comprise $m$ identical equi-spaced apertures with Fourier components

\begin{eqnarray}\label{9a}
\Psi_{\rm Mask}^{\rm (m)}(\phi)=\sum_{q=1}^{m}\Psi_{\rm Mask} \left( \phi + q \frac{2 \pi}{m} \right),
\end{eqnarray}
we obtain OAM sidebands with amplitudes
\begin{eqnarray}\label{9}
\psi_{\delta\ell}^{\rm (m)}=A_{\delta\ell}^{\rm (m)}&=& A_{\delta\ell} \sum_{q=1}^{m} \exp\left(i \frac{q}{m}2\pi\delta\ell \right) \nonumber \\
&=&\left\{\begin{array}{ll}
  m A_{\delta\ell}~&{\rm for}\hspace{5 pt} \delta\ell=N m\\
  0~&{\rm otherwise}\\
  \end{array}\right. ,
\end{eqnarray}
\noindent where $N$ is an integer.  Consequently, in apertures with two-fold symmetry, only every second OAM sideband is present, and in three-fold symmetric apertures only every third.  In our experiments we use masks comprising $m$ hard-edge segments so that only every $m^{\rm th}$ sideband within the sinc envelope is present.

The complex transmission function $\Psi_{\rm Mask}$ may also include phase information.  Specifically we consider the
situation where each of the $m$ hard-edged aperture segments has a definite and non-zero relative phase, $\Phi_q$.  It is
instructive to consider the $m-$fold symmetric composite mask as a superposition of $m$ single-segment apertures, each
giving rise to its own set of OAM sidebands which may constructively or destructive interfere.  This interference between
the individual OAM sidebands constitutes a test of the Fourier relationship between angle and angular momentum.  In
our experiments we investigate one representative case, when the phase of the $m$ segments advances in discrete steps so
that $\Phi_q=\alpha 2 \pi q/m$. The Fourier components and hence OAM sidebands can then be calculated according to
\begin{eqnarray}\label{11}
\psi_{\delta\ell}^{\rm (m,\alpha)}&=&A_{\delta\ell} \sum_{q=1}^{m}\exp\left(i \frac{q}{m}2\pi(\delta\ell+\alpha)\right)   \\
&=&\left\{\begin{array}{ll}
  m A_{\delta\ell} & {\rm for}\hspace{5 pt} \delta\ell=m N-\alpha \\
  A_{\delta\ell} \frac{\exp(i 2\pi(\delta\ell+\alpha))-1}{1-\exp(i 2\pi(\delta\ell+\alpha)/m)}  & {\rm otherwise} \\
\end{array} \right. .\nonumber
\end{eqnarray}
If $\alpha$ is an integer, the central OAM component will be shifted by $\delta \ell=\alpha$, constructive interference will generate OAM sidebands at multiples of $m$, and destructive interference will cancel any other sidebands. If $\alpha$ is not integer, interference between light passing through the different segments will modulate the sidebands.  Note that if $m$ tends to infinity only the central peak at $\delta \ell=\alpha$ remains, turning the mask effectively into a spiral phase plate with optical step height $\alpha \lambda$ \cite{spiral}.  It is worth pointing out that these considerations still hold for pure phase masks by setting $m A_{\delta\ell}=1.$ Such masks generate OAM modes of $\alpha$ modulo $m$, where $m$ is given by the rotational symmetry, and the integer $\alpha$ shifts between the different sets.

\begin{figure}
\includegraphics[width=80mm]{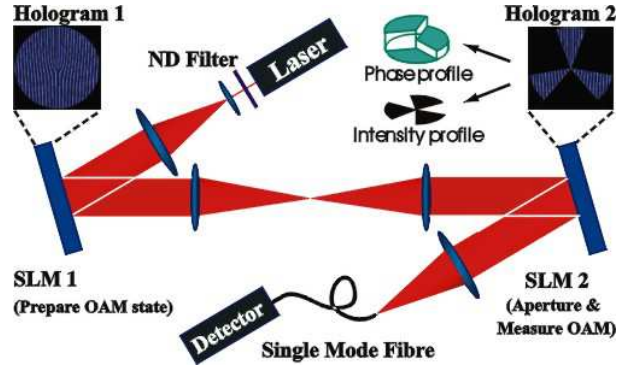}
\caption{Experimental configuration using programmable spatial light modulators to create specific OAM states (SLM1),
aperture them and measure the resulting OAM distribution (SLM2).}
\end{figure}

{\bf Experimental Configuration.}
Within our experimental configuration, see figure 1, we generate a low intensity laser beam in a pure $\ell$-state by transforming a collimated He-Ne laser ($\ell=0$) with a spatial light modulator.  Spatial light modulators act as reconfigurable phase gratings, or holograms, giving control over the complex amplitude of the diffracted beams.  As is standard practice, our modulator is programmed with a diffraction grating containing a fork dislocation to produce a beam with helical phase front in the first diffraction order \cite{Bazhenov}.  A second spatial light modulator is used to analyse the $\ell$-state.
If the index of the analysing hologram is opposite to that of the incoming beam, it makes the helical phase fronts of the incoming beam planar again. A significant fraction of the resulting beam can then be coupled into a single-mode optical fibre.
If the beam and analysing hologram do not match, the diffracted beam has helical phase fronts and therefore no on-axis intensity, resulting in virtually no coupling into the fibre.
To deduce the $\ell$-state, the analysing hologram is switched between various indices whilst monitoring the power transmitted through the fibre. It should be emphasised that the cycling of the hologram index makes the detection process inherently inefficient, where even with perfect optical components, the quantum detection efficiency cannot exceed the reciprocal of the number of different
$\ell$-states to be examined \cite{gibson}.  In principle an amplitude and/or phase mask could be introduced at any position between the two spatial light modulators. However, combining aperture and analysing hologram on a single spatial light modulator eases alignment and improves optical efficiency. We achieve this combination by a modulo $2\pi$ addition of the two holograms.  We measure the light coupled into the single mode fibre with an avalanche photodiode which enables photon counting.
Inserting a neutral density filter immediately after the laser restricts the maximum count rate to less than 100kHz so that at any one time there is on average less than one photon within the apparatus.

To investigate the detailed relation between the angular aperture function and the measured orbital angular momentum states we adopt a family of aperture functions comprising $m$ equi-spaced segments of defined width and relative phase.  For each aperture function the transmitted photons are analysed for the orbital angular momentum states $-18<\ell + \delta\ell<18$.  One complication is that the manufacturing limitations of the spatial light modulators result in deviation from optical flatness by three or four lambda over the full aperture.  This degrades the point-spread function of the diffracted beam
and hence changes the efficiency of the mode coupling into the fibre, spoiling the discrimination between different $\ell$-states.  Therefore, prior to their use within this experiment, we optimise each of the spatial light modulators by applying a hologram of the Zernike terms compansating for  astigmatism to give the smallest point spread function of the diffracted $HG_{00}$ mode.  These correction holograms were added to any of the holograms calculated subsequently.

\begin{figure}
\begin{center}
\includegraphics[width=80mm]{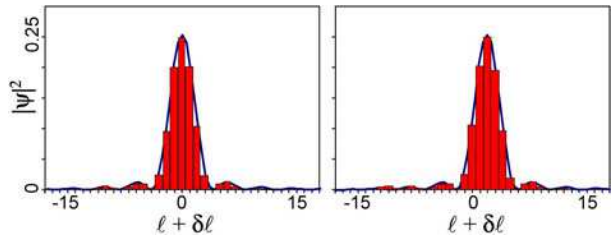}
\caption{The measured (bar) and predicted envelope (line) distribution of OAM sidebands generated from an incident
$\ell=0$ (left) and $\ell=2$ (right) beam after transmission through a hard-edge aperture of angular width $\pi/4$.}
\end{center}
\end{figure}

{\bf Experimental Results.}
For a single hard-edge aperture of uniform phase and width $\Theta $, the resulting integer angular momentum distribution has a sinc function envelope centred on the $\ell$ of the incident mode, as given by (\ref{8}). Figure 2 shows the measured OAM sidebands for a hard-edge aperture of width $\Theta=\pi/4$. We find almost perfect agreement between the observed distribution and that predicted from the Fourier-relation. However, as discussed, a more subtle test of the Fourier-relation is when the aperture function is multi-peaked and when these peaks are offset in phase.  Introducing an aperture comprising two segments of the same width generates an OAM sideband distribution with the same envelope function, but if the Fourier relationship holds true the sidebands can interfere either constructively or destructively depending on the relative phase of the individual components.  Figure 3 compares the angular momentum distribution as predicted from (\ref{11}) with the one observed in the experiment for the case of two (i.e.~$m=2$) diametrically opposed hard-edge apertures, each of angular width $2\pi/9$.  During the experimental sequence their relative phase $\delta\Phi=\Phi_2-\Phi_1$ is varied from $0$ to $2\pi$ ($\alpha$ is varied from 0 to 1). As discussed, the OAM sidebands of the two segments differ in their phase by $(\delta\ell+\alpha)\times\pi$. When $\alpha=0$, the OAM
sidebands with odd $\delta\ell$ interfere destructively.  As the relative phase increases, the light intensity in the odd modes rises at the expense of the even modes until all the even modes disappear when $\delta\Phi=\pi$. At intermediate positions when $\delta\Phi=\pi/2$ or $3 \pi/2$ even and odd sidebands have equal weights. The width of the aperture $\Theta$ changes the width of the sinc distribution but not the underlying interference effects. Figure 4 shows the results for four (i.e.~$m=4$) equi-spaced apertures, at a phase difference of $\alpha\pi/2$. Increasing $\alpha$ from 0 to 4 gives OAM sideband distributions in excellent agreement to that predicted by (\ref{11}).

\begin{figure}
\includegraphics[width=80mm]{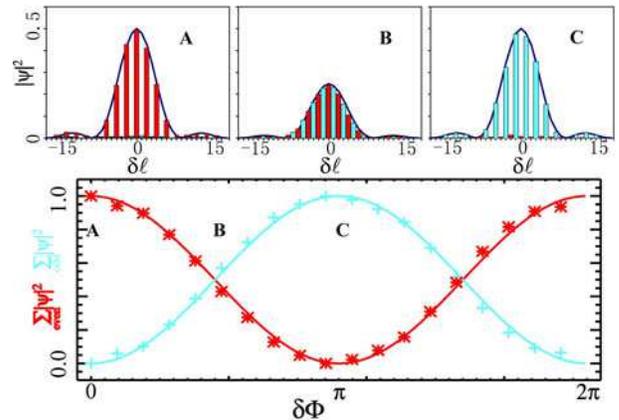}
\caption{ (colour online) Experimental result for two equi-spaced segment apertures of angular width $2\pi/9$. Panels A, B
and C show the observed angular momentum spectra for $\delta\Phi = 0$, $\pi/2$ and $\pi$, respectively, with the predicted
envelope function (solid line) and experimentally measured data (bars). Even sidebands are plotted in red and odd
sidebands in cyan. The main panel shows the sums of coefficients of even (red {\color{my_red} $\blacksquare$}) and odd
(cyan {\color{my_cyan} $\blacksquare$}) OAM sidebands as a function of phase difference, $\delta\Phi$, between the
segments. Solid curves show the predicted variation while experimental data are plotted as stars and crosses.  }
\label{Fig3}
\end{figure}

\begin{figure*}
\includegraphics[width=160mm]{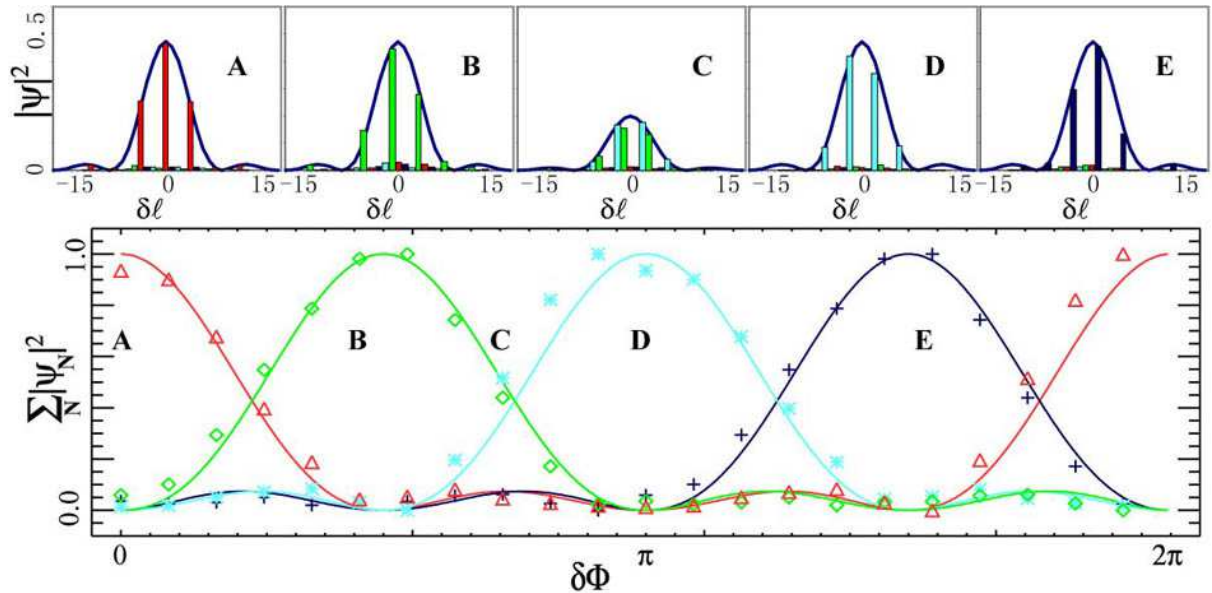}
\caption{(colour online) Experimental result for four equi-spaced apertures of  angular width $2\pi/9$. Sums of
coefficients of all $4N$ (red {\color{my_red} $\blacksquare$}), $4N+1$ (green {\color{my_green} $\blacksquare$}), $4N+2$
(cyan {\color{my_cyan} $\blacksquare$}) and $4N+3$ (blue {\color{my_blue} $\blacksquare$}), where $N$ is an integer,
angular momentum sidebands as a function of phase difference between the two angular apertures. Solid curves show the
predicted variation.} \label{Fig4}
\end{figure*}

{\bf Discussion and Conclusions.}
We have shown that angle and angular momentum states are related as conjugate variables by a Fourier transformation, and that this relationship holds for both amplitude and phase.  Fourier relationships of this type give rise to uncertainty relations between the standard deviations of the conjugate variables.  However, the $2\pi$ cyclic nature of angular measurement raises difficulties in the formulation of an angular uncertainty relation and the definition of a suitable angle operator.  An angle operator should yield results defined within a chosen $2\pi$ radian range \cite{Barnett} .  This approach gives an uncertainty relation which limits the accuracy of possible measurements to  $\Delta \phi\Delta L_{\rm z} \geq(\hbar/2)|1-2 \pi P(\phi_0)|$, where $P(\phi_0)=P(\phi_0+2 \pi)$ is the normalised probability at the limit of the angle range \cite{Franke-Arnold}.  This uncertainty relation may be seen as a consequence of the Fourier-relationship, directly demonstrated in this paper.  Throughout our investigations, we used low light intensities corresponding to single photon flux rates.  Although all our measurements were classical in nature, the results and Fourier-relationship should also hold at the single photon or quantum level.  Furthermore, while demonstrated in the optical regime, the Fourier-relation is expected to be valid for any system having a wave nature including superfluids or BECs.

Hard-edge apertures as used in this investgation, to shape the azimuthal distribution of a light beam, can be used to generate sidebands of the orbital angular momentum or indeed controlled superpositions of particular orbital angular momentum states.  The presence of sidebands may create ambiguities if measured using holographic techniques but such modes are completely compatible with mode sorters based on the rotational symmetry of the modes \cite{modesorter}.

This work was supported by the UK's Engineering and Physical Sciences Research Council, JC and SFA are supported by the Royal Society.

\end{document}